\documentstyle[12pt]{article}
\begin{document}
\begin{normalsize}
\pagestyle{empty}
\baselineskip=15pt
\def\e{\epsilon}
\def\z{\zeta}
\def\h{\eta}
\def\th{\theta}
\def\Th{\Theta}
\def\k{\kappa}
\def\l{\lambda}
\def\L{\Lambda}
\def\m{\mu}
\def\n{\nu}
\def\x{\xi}
\def\X{\Xi}
\def\o{\o}
\def\p{\pi}
\def\P{\Pi}
\def\vp{\varpi}
\def\r{\rho}
\def\vr{\varrho}
\def\s{\sigma}
\def\t{\tau}
\def\f{\phi}
\def\F{\Phi}
\def\vf{\varphi}
\def\c{\chi}
\def\ps{\psi}
\def\Ps{\Psi}
\def\w{\omega}
\def\W{\Omega}
 
\renewcommand{\theequation}{\arabic{section}.\arabic{equation}}
\renewcommand{\thefootnote}{\fnsymbol{footnote}}
 
\def\spc{\makebox[11cm][l]}
\spc{}March 1995
 
\spc{}(revised November 1995)
 
\spc{}TEP-12
 
\spc{}hep-th/9503196
 
$ $\vspace{2cm}
 
\begin{center}
{\Huge {\bf BRS Cohomology of a Bilocal Model}}\footnote[1]{The original 
title is ``On the Dimensions of Space Time''.}\vspace{1cm}
 
  \vspace{1cm}
 
{\bf Takayuki Hori}\footnote[5]{e00353@sinet.ad.jp}\vspace{.5cm}
 
{\it Institute for Physics, Teikyo University}
 
{\it 359 Otsuka, Hachi\=oji-shi, Tokyo, 192-03 Japan}
 
 \vspace{4cm}
 
{\bf Abstract}\vspace{.5cm}
 
\end{center}
 
We present a model in which a gauge symmetry of a field theory 
is intrinsic in the geometry of an extended space time itself.
A consequence  is that the dimension of our space time is 
restricted through the BRS cohomology.
If the Hilbert space is a dense subspace of the space of all 
square integrable $C^{\infty}$ functions, the BRS cohomology 
classes are nontrivial only when the dimension is two or four.\vspace{.5cm} 
 
 \vspace{1.5cm}
\noindent published in {\it Prog. Thoer. Phys. 95 (1996) 803}
\newpage
\pagestyle{plain}
\setcounter{page}{1}
\setcounter{section}{1}
 
\noindent  {\bf 1. Introduction}\vspace{.5cm}
 
The problem why the dimension of our space time is four has not 
been answered so far.  
In the string models the dimensions $D = 26$ or $10$ play a 
special role, but 
any dynamical mechanisms by which the extra dimensions are uniquely
compactified to make our four dimensional world have not been 
known.
In this paper we point out another possibility which restricts, 
if not determines, the dimension of the space time.
A basic idea is to convert the gauge principle of a field theory 
into an intrinsic geometry of the space time.
 
The Minkowski geometry is characterized as the invariant properties 
under the transformations which leave the interval, 
$ds^2 = -\h _{\m \n}dx^{\m}dx^{\n}$, invariant\footnote[2]{We use 
the convention, $\h_{\m \n} = diag(-++...+)$.}.
The trajectories of a particle are the geodesics which minimize 
$-\int ds$.
In the presence of an external electromagnetic field the trajectories
of a particle with mass $m$ and electric charge $e'$ are modified 
from the geodesics to the ones which minimize the action
\begin{eqnarray}
           I_0  = -m\! \int \! ds   + e'\! \int \! A_{\m}(x)dx^{\m}.
\label{I0}
\end{eqnarray}
It is possible to reinterpret the trajectories as  geodesics of an 
extended space spanned by $(x^{\m}, a^{\m})$ where $a$'s are 
fictitious coordinates.
By introducing the internal time $\t$ and the einbein $V$ the 
integrated world length is defined as
\begin{eqnarray}
    I'_0 = \int \! d\t \left[ \frac{1}{2V(\t )}\dot{x} ^{\m}(\t
)\dot{x} _{\m}(\t ) - \frac12 m^2V(\t )   + ea_{\m}(\t )\dot{x} ^{\m}(\t
)\right] ,
\label{I'}
\end{eqnarray}
where dots denote derivatives with respect to $\t$.
Then the trajectories of the charged particle are obtained by 
minimizing the world length $I'_0$ within the hypersurface defined 
by $ea_{\m} = e'A_{\m}(x)$.
(In this interpretation the coupling constant $e$ has the dimension 
of mass square.)
 
The canonical theory is obtained by regarding $\t$ as time.
Introducing canonical variables $p_{\m}$ and $\P$ conjugate 
to $x^{\m}$ and $V$, respectively, we can write the Hamiltonian as
\begin{eqnarray}
          H'  &=& \l L  + \L \P ,\\
         L &=& (p - ea)^2 - m^2
\label{H'}
\end{eqnarray}
where $\l  = V$ and $\L  = \dot{V}$ are arbitrary functions of 
canonical variables.
We get the primary constraint $\P  \sim  0$ and the secondary 
constraint $L  \sim  0$.
The first quantization is achieved by replacing $p_{\m}$ and 
$\P$ by $-i\frac{\partial}{\partial x^{\m}}$ and $-i\frac{\partial}{\partial V}$, 
respectively. The canonical constraint is represented as the wave 
equation $L\Ps(x) =  0$, which coincides, on the hypersurface 
$ea_{\m} = e'A_{\m}(x)$, with the Klein-Gordon equation for a 
charged particle in the external electromagnetic field.
Furthermore, the action of the field theory is given by
\begin{eqnarray}
    {\cal I}' =  \left.\int d^Dxdc\Ps^*Q\Ps \right|_{ea = e'A(x)},
\label{FI'}
\end{eqnarray}
where $Q = cL$ is the BRS operator and $c$ is the ghost variable 
corresponding to the reparametrization gauge of the action $I'_0$.
 
We want to promote the above procedure into the more intrinsic 
geometrical construction.
That is, we start neither with external field  nor constrained 
hypersurface, and treat whole space time spanned by 
$(x^{\m}, a^{\m})$ as the basic space time.
The field theory may contain multiple of fields, hopefully of 
gauge fields, in an expansion of a basic field as
 $\Ps (x, a) = \f(x) + a^{\m}A_{\m}(x) + ...$, and an action like 
eq.(\ref{FI'}), but without the restriction to a hypersurface, may 
describe the dynamics for the fields.
 
The world length defined by eq.(\ref{I'}) is clearly inadequate for 
the above purpose, since it has no information on the gauge 
properties of the electromagnetic field.
For example the physical degrees of freedom should be the spatial 
coordinates, $\vec{x}$, and the transverse {\it polarizations}, 
$\vec{a}_T$, but the reparametrization invariance of $I'_0$ can 
eliminate only one component among $2D$ coordinates.
 
Instead we start with the following world length, which permits 
just the desired degrees of freedom, 
\begin{eqnarray}
     I = \int \! d\t \left[ \frac{1}{2V_1(\t )}\dot{x} ^{\m}(\t )\dot{x}_{\m}
(\t ) 
             +   \frac{1}{2V_2(\t )}\dot{a} ^{\m}(\t )\dot{a} _{\m}(\t ) 
         + 2ea_{\m}(\t )\dot{x} ^{\m}(\t )\right] .
\label{I}
\end{eqnarray}
(the factor 2 in front of $e$ is a convention).
Apart from the above interpretation eq.(\ref{I}) is formally 
regarded as an action for a bilocal particle each ingredient of 
which is described by the coordinates $x^{\m}$ (or $a^{\m}$). 
This model was considered in previous papers \cite{hori1,hori2}. 
There it was shown that the action has a hidden local symmetry of 
$SL(2, R)$, and it is sufficient to eliminate three of $2D$ 
coordinates.
These facts are clearly seen in the canonical theory, but in the 
Lagrangian formalism the gauge symmetry is hidden and quite 
unexpected, since apparently there is only one reparametrization 
parameter if $e \ne  0$.
 
The world length defined by eq.(\ref{I}) is the basic quantity of 
our model, and we will often call it as action in the context of 
the canonical theory which was described in ref.\cite{hori2} as 
a model for a bilocal particle. 
The bilocal particle interpretation, however, cannot be extended 
to the curved space time, since the coordinates $a^{\m}$ would 
not be vector under general coordinate transformations thus $I$ 
cannot be invariant.
But the arguments in the subsequent sections are all in the flat 
space time, and the results are independent of which 
interpretation one chooses.
 
A dynamical theory on the space time who's geometry is determined 
by $I$ should be restricted by the BRS cohomology associated with 
the $SL(2, R)$ symmetry.
There is, however, an arbitrariness of the function space 
which we choose as the basic Hilbert space.
We choose a dense subspace of the space, $\tilde{H}$, of all square 
integrable $C^{\infty}$ functions, which is defined in section 3.
Our main result is that the BRS cohomology classes are 
nontrivial only when $D = 2$ or $4$, {\it i.e.}, the theory 
should be empty otherwise.
When $D = 2$ the nontrivial physical states have only spin one, 
while when $D = 4$ they have only spin zero. 
 
Therefore, a field theory appropriate to the BRS structure of the 
space time may contain  physical gauge fields only when $D = 2$, 
a rather disappointing result.
For $D \ne  2$ the gauge symmetries of a field theory are not 
physical and the gauge fields are all pure gauge.
This means that we must make some extension of the basic space 
time or the Hilbert space in order to get a realistic gauge 
field theory.
Possible extensions are briefly discussed in the final 
section.\vspace{.5cm}
 
In section 2 we give a brief review of the canonical theory in 
the previous paper\cite{hori2}, correcting some minor errors of 
signatures contained in it.
And we define a Hilbert space which is the base of the argument 
on the BRS cohomology.
In section 3 the BRS cohomology is obtained. 
In section 4 a free field theory is defined which is based on 
the BRS structure of the geometry.
Section 5 is devoted to outlooks.
In Appendix, two facts, {\it i.e.}, the linear independence of 
the basis we use, and that the basic function space is dense in 
$\tilde{H}$,  are proved. \vspace{1.5cm}
 
\setcounter{section}{2}
\setcounter{equation}{0}
\noindent {\bf 2. Quantization}\vspace{.5cm}
 
Let us briefly review the result of ref.\cite{hori2}.
We can derive the canonical theory by regarding $\t$ as time.
It is convenient to add a total derivative term, 
$-e\frac{d}{dt}(ax)$, in $I$ to symmetrize $x$'s and $a$'s.
 
The canonical momenta of $x$'s, $a$'s and $V$'s are denoted 
by $p$'s, $b$'s and $\P$'s, respectively.
Then the Hamiltonian is given by
\begin{eqnarray}
    {\cal H} = \l _{1}L_{1} + \l _{-1}L_{-1} + 
\sum_{a=1,2}\L_{a}\Pi _{a},
\label{hamiltonian}
\end{eqnarray}
where $\l$'s and $\L$'s are arbitrary functions of the 
canonical variables, and
\begin{eqnarray}
      L_{1}  =  \frac{i}{4e}(p - ea)^2, \qquad  L_{-1} 
= \frac{i}{4e}(b + ex)^2,
\label{L1}
\end{eqnarray}
(the factor $i$ is a convention, and we omit the space time 
suffices here and hereafter).
The primary constraints are $\Pi _a \sim  0$, the stability 
condition of them along $\t$ development leads to the secondary 
constraints $L_{\pm 1} \sim  0$, and finally the stability of 
the latter requires the tertiary constraint:
\begin{eqnarray}
       L_{0} = \frac{i}{4e}(p -ea)(b + ex) \sim   0.
\end{eqnarray}
 
The constraints $L$'s form a first class algebra and according 
to the Dirac conjecture \cite{Dirac} they may generate a gauge 
symmetry of the action (\ref{I}).
In fact $I$ is invariant under transformations with two local 
parameters $\e _0, \e _1$ and their $\t$ derivatives up to 2 
ranks, $\dot{\e}_1, \dot{\e}_0, \ddot{\e}_0$ \cite{hori2}.
Thus the independent Cauchy data in a time like surface are 
reduced by five, two of which correspond to the einbeins and 
there remains $2D - 3$ physical coordinates as expected.
 
Proceeding to quantum theory, we replace $p$ and $b$ by 
$-i\frac{\partial}{\partial x}$ and $-i\frac{\partial}{\partial a}$, respectively.
The commutators of $L$'s form $SL(2, R)$ algebra:
\begin{eqnarray}
 [L_{n},L_{m}] = (n-m)L_{n+m} + 2(\alpha - \frac{D}{4})\delta _{n+m},  
\qquad (n, m = 0, \pm 1),
\end{eqnarray}
where the central term is caused by the ordering ambiguity in 
$L_0$ defined by
$L_0 =  \frac{i}{4e}(p - ea)(b + ex) - \alpha$.
The BRS operator is 
\begin{eqnarray}
    Q = \sum_{n=0,\pm 1}c_{n}L_n
       -  \frac{1}{2}\sum_{n,m=0,\pm 1}(n-m)c_{n}c_{m}
\frac{\partial}{\partial c_{n+m}},
\end{eqnarray}
where $c_n (n = 0, \pm 1)$ are the ghost variables.
The nilpotency of $Q$ fixes the ordering ambiguity in $L_0$.
In fact we see
\begin{eqnarray}
            Q^2 = 2\left( \alpha  - \frac{D}{4}\right) c_1c_{-1},
\label{nilpo}
\end{eqnarray}
which requires $\alpha = \frac{D}{4}$, so we have
\begin{eqnarray}
       L_0 =  \frac{i}{4e}(-i\frac{\partial}{\partial x} - 
ea)(-i\frac{\partial}{\partial a} + ex) - \frac{D}{4}.
\label{L0=}
\end{eqnarray}
(Note $L_0 = \frac{i}{8e}\{ p - ea, b + ex\} $.)
 
Next let us define the generators of the kinematic symmetry:
\begin{eqnarray}
              \tilde{p}  = p  + ea,  \qquad  \tilde{b}  = b  - ex,
\label{pt}
\end{eqnarray}
\begin{eqnarray}
             M_{\m \n} = x_{[\m}p_{\n ]} + a_{[\m}b_{\n ]}, 
\label{Lorentz}
\end{eqnarray}
These generators satisfy the ordinary commutation relations 
except that $[\tilde{p} _{\m}, \tilde{b} _{\n}] = 2ei\h _{\m \n}$ which is 
interpreted as an uncertainty relation.
It is important to note that the kinematic generators all 
commute with $L_n, (n = 0, \pm 1)$:
\begin{eqnarray}
    [\tilde{p} _{\mu}, L_{0,\pm 1}] =  [\tilde{b} _{\mu}, L_{0,\pm 1}] 
= [M_{\mu \nu}, L_{0,\pm 1}] = 0.
\label{Kin-L}
\end{eqnarray}
 
There is the unique ground state, $|0\rangle$, which is annihilated 
by $L_{-1}$ and Lorentz invariant and has vanishing 
momentum\footnote[2]{The state $|0\rangle$ here is different from 
that defined in ref.\cite{hori2}. The latter is an eigenstate 
of {\it total} momentum, $\tilde{p}  + \tilde{b}$.},
\begin{eqnarray}
                  |0\rangle   &=&  e^{-ieax} ,
\label{|0>}\\
         M_{\m \n}|0\rangle  &=& \tilde{p}_{\m}|0\rangle   = L_{-1}|0\rangle  =  0.
\label{M=p=L1=0}
\end{eqnarray}
We define the state with momentum $k$ by
\begin{eqnarray}
           |k\rangle   = e^{-\frac{i}{4e}\tilde{b} k} |0\rangle  = e^{ixk}|0\rangle .
\label{|k>}
\end{eqnarray}
 
According to the Dirac prescription the first quantization 
would be achieved by requiring the wave equations, 
$L_{0, \pm 1}\Ps(x, a) = 0$.
Here $\Ps$ would be any function which is square integrable 
and differentiable to an arbitrary order.
Let us denote the set of all such functions by $\tilde{H}$.
%
We assume, however, that the Hilbert space is not the whole space $\tilde{H}$ but a dense subspace of $\tilde{H}$.
One can prove the existence of the subspace, $H_1$, which contains an arbitrary momentum and spin states, and a function in $H_1$ is general enough as $H_1$ is dense in the whole function space $\tilde{H}$.
We find that the following functions span the dense subspace $H_1$ in $\tilde{H}$:
\begin{eqnarray}
  u_{nJj}(k)  =  L^n_1e^{-\frac{i}{4e}\tilde{b} k} F_{Jj}(a)|0\rangle  \qquad (n, J, j = 0,1,...),
\label{base}
\end{eqnarray}
where $F_{Jj}(a)$ are the harmonic polynomials of $a$ with 
homogeneous order $J$, which satisfy $
        {\rule[-1mm]{0.1mm}{4mm}
         \rule[3mm]{4mm}{0.1mm} 
         \hspace{-4mm}
         \rule[-1mm]{4mm}{0.1mm} 
         \rule[-1mm]{0.1mm}{4mm}
         \hspace{1mm}}_aF_{Jj} = 0$, and 
$j$ varies from $1$ to $(2J + D - 2)(J + D - 3)!/J!(D - 2)!$ 
\cite{Take}.
We denote by $H_1$ the set of all functions which can be 
written as  linear combinations (and integrations over $k$) of 
$u_{nJj}(k)$, with vanishing coefficients except a finite 
number of ones.
 
In Appendix, we show that $H_1$  is dense in $\tilde{H}$, and that a 
finite number of element in $\{ u_{nJj}(k)\} $ are linearly independent.
Thus the function space $H_1$ satisfies all the requirements 
mentioned before, and we choose it as the basic Hilbert space.
A merit of our basis (\ref{base}) is that they belong to a 
representation of the $SL(2, R)$ as expressed in eqs.(\ref{A5}), 
(\ref{A6}) and (\ref{A7}) in Appendix, and a calculation to obtain 
the BRS cohomology becomes algebraic. \vspace{1.5cm}
 
\setcounter{section}{3}
\setcounter{equation}{0}
\noindent {\bf 3. BRS cohomology}\vspace{.5cm}
 
Let us obtain the BRS cohomology classes of the first quantized 
system.
Our task is to obtain all classes of functions $\Ps$'s in $H_1$, 
which satisfy the Kugo-Ojima(KO) condition \cite{KO},
\begin{eqnarray}
 Q|\Ps \rangle  = 0,
\label{KO}
\end{eqnarray}
{\it and} are not written as $|\Ps \rangle   = Q|\c \rangle$ for some 
$|\c \rangle$ in $H_1$.
Since the ghost variables are Grassmann odd we can divide $H_1$ 
according to the ghost numbers, $N_g$, varying from 0 to 3.
We can search such functions separately in each sector because 
$Q$ has ghost number $1$ and the KO condition does not mix 
sectors with different ghost numbers.
 
 
\noindent (1) $N_g  = 0$: Let us write a function $\Ps^{(0)}\in H_1$, as
\begin{eqnarray}
      \Ps^{(0)}  = \sum_{n = 0}^{\infty}\sum_{J = 0}^{\infty}\sum_{j}\int_k\alpha_{nJj}(k)u_{nJj}(k).
\label{Ps0}
\end{eqnarray}
In this sector the KO condition (\ref{KO}) amounts to 
$L_{n}\Ps  = 0$ for $n = 0, \pm 1$.
Although the subscript of the summation in eq.(\ref{Ps0}) extend to
infinity, only a finite number of the coefficients are nonvanishing.
In Appendix we show an arbitrary finite subset of $\{ u_{nJj}(k)\} $ is 
linearly independent.
Therefore, using only $L_1\Ps = 0$, we see, without no subtleties 
on divergence or infinite summations, that  all $\alpha$'s vanishes.
Thus in this sector of $H_1$ there is no physical state.
 
\noindent (2) $N_g = 1$: In this sector we can write an arbitrary state as
\begin{eqnarray}
 \Ps^{(1)}  = \sum_{n = 0}^{\infty}\sum_{J = 0}^{\infty}(\alpha_{nJ}c_0 + \beta_{nJ}c_1 
+ \gamma_{nJ}c_{-1})u_{nJ},
\label{Ps1}
\end{eqnarray}
where $\alpha$'s, $\beta$'s and $\gamma$'s are numerical coefficients.
Integrations over $k$ are implicitly assumed and we suppress the 
$k$-dependence in the expressions, since all $L$'s commute with 
$\tilde{b}$ and they play no role in the present argument ($j$ dependence 
is also suppressed). 
 
The KO condition requires the following equations to the 
coefficients, for $n \ge 1$ :
\begin{eqnarray}
                 \left( \begin{array}{ccc}  2 & (n + 1)(n + \frac{D}{2} + J )  & -1 \\
                         1 & n + \frac{D}{4} + \frac{J}{2}   &  0 \\
              n(n - 1 + \frac{D}{2} + J ) & 0    & n + \frac{D}{4} + \frac{J}{2}
                 \end{array} \right)
                 \left( \begin{array}{c}  \alpha _{nJ} \\
                        \beta _{n + 1 J}\\
                        \gamma _{n - 1 J}
                 \end{array} \right)   = 0,
\label{KO1-1} 
\end{eqnarray}
and 
\begin{eqnarray}
                \left( \begin{array}{cc} 2 & \frac{D}{2} + J \\
                      1 & \frac{D}{4} + \frac{J}{2}
                \end{array} \right)
                \left( \begin{array}{c} \alpha_{0J}\\
                      \beta_{1J}
                \end{array} \right)    = 0,
\label{KO1-2}
\end{eqnarray}
\begin{eqnarray}
                 \left( 1  - \frac{D}{4} - \frac{J}{2}\right) \beta_{0J}  = 0.
\label{KO1-3}
\end{eqnarray}   
These equations are a consequence of the (finite) linear 
independence of our basis (\ref{base}).
Since the matrices appearing in eqs.(\ref{KO1-1}) and (\ref{KO1-2}) 
have vanishing determinants, we get the following solution
\begin{eqnarray}
  \Ps ^{(1)} &=& \sum_{J = 0}^{\infty}\Big[\beta_{0J}c_1u_{0J} 
+ \beta_{1J}(-(\frac{D}{4} + \frac{J}{2} )c_0u_{0J} + c_1u_{1J})\nonumber \\
    &+& \sum_{n = 1}^{\infty}\frac{\alpha_{nJ}}{n + \frac{D}{4} + \frac{J}{2}}\left( (n + \frac{D}{4} + \frac{J}{2} )c_0u_{nJ}  - 
c_1u_{n + 1 J} - n(n - 1 + \frac{D}{2} + J )c_{-1}u_{-1 J}\right)\Big] .
\label{sol-1}\nonumber \\
{}
\end{eqnarray}
A part of expression in r.h.s of eq.(\ref{sol-1}) may be written 
as a BRS trivial form.
Since the BRS trivial quantities in the present sector have the 
ghost number 1, only candidates are of the form  $Qu_{nJ}$:
\begin{eqnarray}
      Qu_{nJ} &=&  -(n + \frac{D}{4} + \frac{J}{2} )c_0u_{nJ}  + c_1u_{n + 1 J} + 
n(n - 1 + \frac{D}{2} + J )c_{-1}u_{n - 1 J} \qquad (n\ge  1),
\label{QA1}\nonumber \\
{}\\
      Qu_{0J} &=& -(\frac{D}{4} + \frac{J}{2} )c_0u_{0J} + c_1u_{1J}.
\label{QA2}
\end{eqnarray}
Comparing eq.(\ref{sol-1}) with eqs.(\ref{QA1}, \ref{QA2}) we find
\begin{eqnarray}
 \Ps^{(1)}  = \sum_{J = 0}^{\infty}\beta_{0J}c_1u_{0J}  
- Q\left(\sum_{J = 0}^{\infty}\left(\beta_{1J}u_{0J} 
+ \sum_{n = 0}^{\infty}\frac{\alpha _nu_n}{n + \frac{D}{4} + \frac{J}{2}}\right)\right).
\label{Sol-1}
\end{eqnarray}
Furthermore, by eq.(\ref{KO1-3}) we find that if 
$(D, J) \ne  (2, 1), (4, 0)$ then $\beta_0  = 0$.
Thus if $D \ne  2, 4$, the solution to the KO condition is BRS 
trivial, while if $D = 2 (J = 1)$ or $D = 4 (J = 0)$ we get the 
BRS nontrivial 
physical states, $c_1u_{0J}$.
 
\noindent (3) $N_g = 2$: In this sector we have the expansion
\begin{eqnarray}
   \Ps^{(2)}  = \sum_{n = 0}^{\infty}\sum_{J = 0}^{\infty}(\alpha'_{nJ}c_0c_{-1}  + \beta'_{nJ}c_0c_1 + 
\gamma'_{nJ}c_1c_{-1})u_{nJ}.
\label{Ps2}
\end{eqnarray}
The solution to the KO condition is similarly obtained as before:
\begin{eqnarray}
    \Ps^{(2)} &=& \sum_{J = 0}^{\infty}\Big[\beta'_{0J}c_0c_1u_{0J}  
+ \beta'_{1J}(c_0c_1u_{1J} + 2c_1c_{-1}u_{0J})\nonumber \\
         & &  + \sum_{n = 1}^{\infty}\gamma'_n\left(c_1c_{-1}u_{nJ}  -  
(n + \frac{D}{4} + \frac{J}{2} )c_0c_{-1}u_{n - 1 J}\right) \nonumber \\
         & &  + \sum_{n = 2}^{\infty}\beta'_n\left(c_0c_1u_{nJ}  +  
n(n - 1 + \frac{D}{2} + J )c_0c_{-1}u_{n - 2 J}\right)\Big],
\label{sol-2}
\end{eqnarray}
where the coefficients in eq.(\ref{sol-2}) are all arbitrary.
The list of the BRS trivial states in the present sector is as 
follows:
\begin{eqnarray}
 &&          \left( \begin{array}{c} Qc_0u_{nJ}\\
                   Qc_1u_{n + 1 J}\\
                   Qc_{-1}u_{n - 1 J}
             \end{array} \right)    = \nonumber \\
 && \quad     \left( \begin{array}{ccc}  2 & 1 & n(n - 1 + \frac{D}{2} + J )\\
                     (n + 1)(n + \frac{D}{2} + J )  & n + \frac{D}{4} + \frac{J}{2}  & 0 \\
                     -1  & 0 & n+ \frac{D}{4} + \frac{J}{2}
             \end{array} \right)
             \left( \begin{array}{c}   c_{-1}c_1u_{nJ}\\
                     c_1c_0u_{n + 1 J}\\
                     c_{-1}c_0u_{n - 1 J}
             \end{array} \right)  ,
\label{KO2-1}
\end{eqnarray}
\begin{eqnarray}
             \left( \begin{array}{c}  Qc_0u_{0J}\\
                    Qc_1u_{1J}
             \end{array} \right)   = 
             \left( \begin{array}{cc}  2 & 1\\
                    \frac{D}{2} + J  & \frac{D}{4} + \frac{J}{2} 
             \end{array} \right)
             \left( \begin{array}{c}   c_{-1}c_1u_{0J}\\
                     c_1c_0u_{1J}
             \end{array} \right)  ,
\label{KO2-2}
\end{eqnarray}
\begin{eqnarray}
        Qc_1u_{0J} = \left( 1 - \frac{D}{4} - \frac{J}{2} \right) c_0c_1u_{0J}.
\label{KO2-3}
\end{eqnarray}
Note the matrices appeared in eqs.(\ref{KO2-1}, \ref{KO2-2}) are 
transposition of those in eqs.(\ref{KO1-1}, \ref{KO1-2}), hence 
have vanishing determinants and are not invertible.
Thus only specific combinations of $c_ac_bu_{nJ}$ are BRS trivial.
We find all terms, except the first one, in r.h.s. of 
eq.(\ref{sol-2}) is BRS trivial,
\begin{eqnarray}
 \Ps ^{(2)} &=& \sum_{J = 0}^{\infty}\beta'_0c_0c_1u_{0J}\nonumber \\
& &  + Q\Big(\sum_{J = 0}^{\infty}\Big\{-\beta'_1c_0u_{0J}  
+ \sum_{n = 1}^{\infty}\gamma'_nc_{-1}u_{n - 1 J}\nonumber \\
& & \qquad - \sum_{n = 2}^{\infty}\frac{\beta'_n}{n - 1 + \frac{D}{4} + \frac{J}{2}}(c_1u_{n + 1 J} 
+ (n + 1)(n + 
\frac{D}{2} + J )c_{-1}u_{n - 1 J})\Big\}\Big) . 
\label{Sol-2}
\end{eqnarray}
Furthermore  we find that if $D \ne  2, 4$ then the factor 
$(1 - \frac{D}{4} - \frac{J}{2} )$ in eq.(\ref{KO2-3}) is invertible, and the first 
term in r.h.s. of eq.(\ref{Sol-2}) is also BRS trivial.
Thus if $D \ne  2, 4$, the solution to the KO condition is BRS 
trivial, while if $D = 2 (J = 1)$ or $D = 4 (J = 0)$ we get the 
BRS nontrivial physical states, $c_0c_1u_{0J}$.
 
\noindent (4) $N_g = 3$: In this sector there exists no BRS nontrivial 
physical states, since
\begin{eqnarray}
    Qc_1c_{-1}u_{nJ} = -\left(n + \frac{D}{4} + \frac{J}{2} \right)c_0c_1c_{-1}u_{nJ}.
\end{eqnarray}
This is the final possibility.
   
We conclude from the above arguments for the four cases that if 
$D \ne  2, 4$ then the BRS cohomology of our system is trivial, 
while if $D = 2 (J = 1)$ or $D = 4 (J = 0)$ then it contains two 
classes which are BRS equivalent to $c_1u_{0Jj}(k)$ or 
$c_0c_1u_{0Jj}(k)$. 
In the framework of the first quantization, the mass spectrum of the physical states of the bilocal system is continuous.
This is because we have no interactions, in the ordinary sense, between the two particles, but the two particles are correlated through the internal $SL(2, R)$ local symmetry generalizing the reparametrization of the world lines.
A discrete mass spectrum, if exists, might be obtained after introducing a possible interaction between the bilocal particles in the framework of the second quantization.
In the free field theory defined in the next section we see such an indication of the discrete mass spectrum. \vspace{1.5cm}
 
\setcounter{section}{4}
\setcounter{equation}{0}
\noindent {\bf 4. Field theory}\vspace{.5cm}
 
Let us examine how the inherent BRS structure of the basic space 
time $M^{2D}$, spanned by $(x^{\m}, a^{\m})$, is connected with a 
gauge symmetry of a field theory on $M^{D}$.
As an example we present here the free field theory.
 
A field $\Ps$ is a function of $x$'s, $a$'s and $c$'s, and we 
assume $\Ps$ is Grassmann odd.
The action is defined by
\begin{eqnarray}
          {\cal I}  = \int\!d^D\!xd^D\!adc_0dc_1dc_{-1}\Ps^*Q\Ps,
\label{action}
\end{eqnarray}
which is invariant under the following gauge transformation,
\begin{eqnarray}
         \delta \Ps  = Q\L,
\label{gaugeTr}
\end{eqnarray}
where $\L$ is an arbitrary function of $x$'s, $a$'s and $c$'s, 
which is Grassmann even.
Since the BRS operator has the ghost number one, the only 
component fields of $\Ps$ with a non-zero ghost number are 
subjected to the gauge transformations.
The component fields of $N_g = 1$ sector have kinetic terms 
in ${\cal I}$ among themselves, and the fields of $N_g = 0$ or $2$ 
have mixed kinetic terms.
 
In general, however, the gauge symmetry is not a physical one.
Since vector fields appear as component of $a^{\m}$ in $\Ps$ and 
belong to $J = 1$ sector (and scalars belong to $J = 0$ sector), 
there exists physical vector fields only when $D = 2$ as was shown 
in the previous section.
For $D \ne  2$ the component fields of $J = 1$ sector are expressed 
as $Q\c$ for some $\c$.
In other words the gauge fields are pure gauge, in the sense of 
eq.(\ref{gaugeTr}), in all cases except $D = 2$.
 
Finally let us write the action ${\cal I}$ as an integration over 
$x$'s of ordinary fields on $M^{D}$.
For that purpose let us expand $\Ps$ as
\begin{eqnarray}
      \Ps  &=& N^{-1/2}_{\e} e^{-\frac12 \e a^2}\sum_{\alpha = 1, \pm 1}c_{\alpha}(\f_{\alpha}(x) + 
A_{\m \alpha}(x)a^{\m} + ... )|0\rangle  + ...,  
\label{Ps=f+}\\
   N_{\e}  &=& \int \!d^D\!ae^{-\e a^2}   
= \quad const.\e^{-\frac{D}{2}},
\end{eqnarray}
where the factor $e^{-\frac12 \e a^2}$ is introduced for making integrals 
convergent.
The explicit form of the action (in the sector with ghost 
number one) is expressed as
\begin{eqnarray}
 {\cal I}  &=&  \int \!d^D\!x \Big\{
 \frac{i}{4e}\Big[\f^*_{0}\Big(
        {\rule[-1mm]{0.1mm}{4mm}
         \rule[3mm]{4mm}{0.1mm} 
         \hspace{-4mm}
         \rule[-1mm]{4mm}{0.1mm} 
         \rule[-1mm]{0.1mm}{4mm}
         \hspace{1mm}}
  - \frac{2e^2D}{\e}\Big)\f_{-1} - 
\f^*_{-1}\Big(
        {\rule[-1mm]{0.1mm}{4mm}
         \rule[3mm]{4mm}{0.1mm} 
         \hspace{-4mm}
         \rule[-1mm]{4mm}{0.1mm} 
         \rule[-1mm]{0.1mm}{4mm}
         \hspace{1mm}}
 - \frac{2e^2D}{\e}\Big)\f_0\Big] \nonumber \\
&& \qquad + \frac{i}{8e\e}\Big[A^{\m *}_{-1}\Big(
        {\rule[-1mm]{0.1mm}{4mm}
         \rule[3mm]{4mm}{0.1mm} 
         \hspace{-4mm}
         \rule[-1mm]{4mm}{0.1mm} 
         \rule[-1mm]{0.1mm}{4mm}
         \hspace{1mm}}
  - 
\frac{2(D + 2)e^2}{\e}\Big)A_{0\m} - A^{\m *}_{0}\Big(
        {\rule[-1mm]{0.1mm}{4mm}
         \rule[3mm]{4mm}{0.1mm} 
         \hspace{-4mm}
         \rule[-1mm]{4mm}{0.1mm} 
         \rule[-1mm]{0.1mm}{4mm}
         \hspace{1mm}}
  - 
\frac{2(D + 2)e^2}{\e}\Big)A_{-1\m}\Big] \nonumber \\
 && \qquad + \frac{1}{2\e}(\f^*_{0}\partial_{\m}A^{\m}_{-1} - 
A^{\m *}_{-1}\partial_{\m}\f_0 + A^{\m *}_{0}\partial_{\m}\f_{-1} - 
\f^*_{-1}\partial_{\m}A^{\m}_{0}) \nonumber \\ 
&& \qquad + \frac{i}{8e}(A^{\m *}_{1}\partial_{\m}\f_{-1} 
 + \f^*_{-1}\partial_{\m}A^{\m}_1 
- A^{\m *}_{-1}\partial_{\m}\f_1 
 - \f^*_{1}\partial_{\m}A^{\m}_{-1})\nonumber \\ 
&& \qquad + 2\f^*_0\f_0 - \f^*_{-1}\f_1 - \f^*_1\f_{-1} + 
\frac{i\e D}{32}(\f^*_{1}\f_{0} - \f^*_{0}\f_{1}) \nonumber \\
&& \qquad - \frac{1}{2\e}(A^{\m *}_{-1}A_{1\m} + A^{\m *}_1A_{-1\m} + 
A^{\m *}_0A_{0\m}) + \frac{i(D + 10)}{64e}(A^*_{1\m}A^{\m}_{0} - 
A^*_{0\m}A^{\m}_{1})
\Big\} .
\end{eqnarray}
Writing the gauge parameter as
\begin{eqnarray}
                \L(x, a)  = N^{-1/2}_{\e} e^{-\frac12 \e a^2}(\L(x) + 
\L_{\m}(x)a^{\m} + ...)|0\rangle  + ...,
\label{Lambda}
\end{eqnarray}
the gauge transformation is expressed as
\begin{eqnarray}
           \delta\f_1  = -\frac{i}{4e}
        {\rule[-1mm]{0.1mm}{4mm}
         \rule[3mm]{4mm}{0.1mm} 
         \hspace{-4mm}
         \rule[-1mm]{4mm}{0.1mm} 
         \rule[-1mm]{0.1mm}{4mm}
         \hspace{1mm}}\L , \qquad
           \delta\f_0  = -\frac{D}{4}\L - \frac{i}{4e}\partial_{\m}\L^{\m},\qquad
           \delta\f_{-1}  = 0 ,
\end{eqnarray}
\begin{eqnarray}
      \delta A_{1\m}  =  -\partial_{\m}\L  - \frac{i}{4e}
        {\rule[-1mm]{0.1mm}{4mm}
         \rule[3mm]{4mm}{0.1mm} 
         \hspace{-4mm}
         \rule[-1mm]{4mm}{0.1mm} 
         \rule[-1mm]{0.1mm}{4mm}
         \hspace{1mm}}\L_{\m},\qquad
      \delta A_{0\m}  =  -\frac{i}{4e}\e\partial_{\m}\L  - \frac{D + 2}{4}\L_{\m},\qquad
      \delta A_{-1\m}  =  \frac{i}{4e}\e\L_{\m}.
\label{gaugeTrC}
\end{eqnarray}
In the above formulas, the divergent factor $\frac{1}{\e}$ can be 
absorbed into the coupling constant $e$ and renormalization 
factors of fields. \vspace{1.5cm}
 
\setcounter{section}{5}
\setcounter{equation}{0}
\newpage
\noindent {\bf 5. Outlooks}\vspace{.5cm}
 
We have shown that the BRS cohomology classes in the Hilbert space 
$H_1$ are nontrivial only when $D = 2$ or $4$.
In the case $D = 2$ the nontrivial physical states are 
$c_1u_{01j}(k)$ and $c_0c_1u_{01j}(k)$, {\it i.e.}, momentum 
eigenstates with spin one, while in the case $D = 4$ they are 
$c_1u_{00}(k)$ and $c_0c_1u_{00}(k)$ which are spin zero states.
In particular the ground state with the vacuum quantum number is 
physical only when $D = 4$ (see \cite{hori2}).
 
Our original hope was that the BRS structure of the basic space 
time would be translated into a gauge symmetry of the 
corresponding field theory.
But it turned out that a possible gauge field belong to a BRS 
trivial sector and is pure gauge except $D = 2$.
Hence a realistic model would be obtained by some modifications or 
extensions of the present one.
 
First possibility is to extend the fictitious dimensions, 
$a^{\m}$, to a multiple of ones, $a^{\m}_i, (i = 1, ..., N)$.
A preliminary investigation shows that the symmetry of the world 
length is enlarged to $Sp(\frac{N + 1}{2})$ for odd $N$, and the 
physical degrees of freedom of the coordinates reduces by 
$\frac12 (N + 1)(N + 2)$.
In order to maintain at least one physical component of the 
coordinates we have the inequality, $\frac12 (N + 2) < D$, which 
means $N \le  5$ for $D = 4$.
This may provide us the more abundant structure, though it is not 
certain whether one of them includes a realistic theory.
A second possibility is to supersymmetrize the model, which may 
introduce fermionic fields.
It is tempting to seek for a supersymmetric world length.
 
Apart from the above directions it is conceivable to enlarge the 
Hilbert space.
For example we may add base functions which are created from a 
vacuum annihilated by $L_1$.
But it turns out, by a similar argument as in section 3, that the 
result on the dimensionality would not be altered from that 
obtained in the present analysis.
 
The model presented here can be regarded as a particular mode of 
the string models.
If one put, {\it e.g.},
\begin{eqnarray}
   X(\t , \s) &=& x(\t) + (2 - 9\s  + 10\s^3)a(\t),\\
 g_{nm}  &=& \left( \begin{array}{cc}  0 & e\\
    e &  \frac{-7 + 12\s}{10V_1(\t)} + \frac{14(1 - \s)}{15V_2(\t)}  
                \end{array} \right)  ,
\end{eqnarray}
the world length (\ref{I}) is written as the world area 
$I = \frac{1}{e}\int 
d{\t}\int_{0}^{1}d\s\sqrt{-g}g^{nm}\partial_{n}X\partial_{m}X$.
The correspondence is not so beautiful and we may not expect any 
intrinsic connections between the two models.
But it is impressive that both models require critical dimensions, 
though in quite different mechanisms.
\newpage

 
 
 
\setcounter{section}{1}
\setcounter{equation}{0}
 
\renewcommand{\theequation}{\Alph{section}.\arabic{equation}}
 
\noindent  {\bf Appendix}\vspace{.5cm}
 
In this Appendix we show the linear independence of the basis 
$\{ u_{nJj}(k)\} $ and that the  space $H_1$  is dense in $\tilde{H}$, the set 
of all square integrable $C^{\infty}$ functions.
Since an element in $H_1$ is a linear combination of 
$u_{nJj}(k)$  with a {\it finite} number of nonvanishing 
coefficients, we show that arbitrary {\it finite} subset 
of $\{ u_{nJj}(k)\} $ is linearly independent.
 
Before proceeding to the proof let us recapitulate useful 
relations.
{}From the commutators of $L_a (a = 0, \pm 1)$ we get 
\begin{eqnarray}
 [L_0, L^n_{\pm 1}] &=& \mp nL^n_{\pm 1},  \label{A1}\\
 {[}L_{\mp 1}, L^n_{\pm 1}] &=& L^{n - 1}_{\pm 1}n(n - 1 \mp 2L_0).
\label{A2}
\end{eqnarray}
For arbitrary polynomials $G(a)$ of $a$, we have
\begin{eqnarray}
   L_{-1}G(a)|0\rangle  &=& -\frac{i}{4e}
        {\rule[-1mm]{0.1mm}{4mm}
         \rule[3mm]{4mm}{0.1mm} 
         \hspace{-4mm}
         \rule[-1mm]{4mm}{0.1mm} 
         \rule[-1mm]{0.1mm}{4mm}}_aG(a)|0\rangle , \label{A3} 
\end{eqnarray}
and for arbitrary homogeneous polynomials $G_J(a)$ of order $J$, 
we have
\begin{eqnarray}
 L_{0}G_J(a)|0\rangle  &=& -\left( \frac{D}{4} 
+ \frac{J}{2}\right) G_J(a)|0\rangle . 
\label{A4} 
\end{eqnarray}
Applying these relations to the harmonic polynomials, $F_{Jj}$, 
satisfying $
        {\rule[-1mm]{0.1mm}{4mm}
         \rule[3mm]{4mm}{0.1mm} 
         \hspace{-4mm}
         \rule[-1mm]{4mm}{0.1mm} 
         \rule[-1mm]{0.1mm}{4mm}
         \hspace{1mm}}_aF_{Jj} = 0$, we get
\begin{eqnarray}
  L_1u_{nJj}(k) &=& u_{n + 1 Jj}(k). \label{A5}\\
  L_{-1}u_{nJj}(k) &=& n\left( n - 1 + \frac{D}{2} + 
J\right)u_{n-1 Jj}(k), \label{A6}\\
  L_{0}u_{nJj}(k) &=& -\left( n + 
\frac{D}{4} + \frac{J}{2}\right)u_{nJj}(k). \label{A7}
\end{eqnarray}
 
Now let us give the proof of the (finite) linear independence of 
$\{ u_{nJj}(k)\} $.
 
\noindent (1) Linear independence.
 
We prove that a finite subset of $\{ u_{nJj}(k)\} $ is linearly independent.
Suppose that
\begin{eqnarray}
     \sum_{n = 0}^{n_0}\sum_{Jj}\int_{k}\alpha_{nJj}(k)u_{nJj}(k)  = 0,
\label{AA1}
\end{eqnarray}
(where $\int_{k}  \equiv  \int \frac{d^Dk}{(2\pi^D}$).
Multiplying $L^{n_0}_{-1}$ to eq.(\ref{AA1}) and using 
eqs.(\ref{A2}) and (\ref{A6}) we see all terms except $n = n_0$ 
vanish and get
\begin{eqnarray}
          \sum_{Jj}\int_{k}\alpha_{n_0Jj}(k)u_{0Jj}(k)  = 0.
\label{AA2}
\end{eqnarray}
Thus we get
\begin{eqnarray}
          \sum_{Jj}\hat{\alpha}_{n_0Jj}(x)F_{Jj}(a)  = 0,
\label{AA3}
\end{eqnarray}
where
\begin{eqnarray}
          \hat{\alpha}_{nJj}(x)  = \int_{k}e^{ikx}\alpha_{nJj}(k).
\label{AA4}
\end{eqnarray}
By the linear independence of the harmonic polynomials we see 
all $\hat{\alpha}_{n_0}$'s vanish, hence also all $\alpha_{n_0}$'s vanish.
Then multiplying $L^{n_0 - 1}_{-1}$ to eq.(\ref{AA1}) we 
get $\alpha_{n_0-1 Jj}(k) = 0$ in the same manner, and so on.
Thus all $\alpha$'s vanish, which completes the proof for the 
finite linear independence of $\{ u_{nJj}(k)\} $.\vspace{.5cm}
 
\noindent (2)  $\bar{H_1} = \tilde{H}$.
 
We show that $H_1$ is dense in $\tilde{H}$, {\it i.e.}, for an 
arbitrary element $f$ in $\tilde{H}$ there exists a sequence $
\{f_{N}\}$ in $H_1$, which converges to $f$ in the limit 
$N \rightarrow  \infty$.
 
Suppose an arbitrary function $f\in \tilde{H}$ is given.
Since $f(x, a)e^{ie(x - a)a}\in \tilde{H}$ it is Fourier expandable:
\begin{eqnarray}
 f(x, a)e^{ie(x - a)a} = \int_{k}\int_{k'}e^{ikx 
+ ik'a}\hat{f}(k, k').
\label{AB1}
\end{eqnarray}
If we define
\begin{eqnarray}
   f_{N}(x, a)  = \sum_{n = 0}^{N}\int_{k}\int_{k'}\hat{f}(k, 
-k')e^{i(k - k')x}\frac{(-L_1)^n}{n!}e^{-\frac{i}{2e}(\tilde{b} k' + 
\frac12 k'^2)}|0\rangle ,
\label{AB2}
\end{eqnarray}
then, using eqs.(\ref{AB1}) and 
\begin{eqnarray}
       L_1|0\rangle  = -iea^2|0\rangle , \qquad  e^{-\frac{i}{2e}(\tilde{b} k' + 
\frac12 k'^2)}e^{iea^2}|0\rangle  = e^{ik(x - a)}e^{iea^2}|0\rangle,
\label{AB3}
\end{eqnarray}
we see that
\begin{eqnarray}
         f(x, a)  =  \lim_{N\to\infty}f_{N}(x, a).
\label{AB5}
\end{eqnarray}
Thus, if $f_{N}\in H_1$ we see $\bar{H} = \tilde{H}$.
 
By moving $e^{i(k - k')x}$ in eq.(\ref{AB2}) to the far right 
until to hit  $|0\rangle$ we get a polynomial of $k, k'$ and $p'$, 
where $p'  \equiv  p - ea$.
Using $p'|0\rangle  = -2ea|0\rangle$ we get a polynomial of $a$ 
multiplied by $e^{i(k - k')x}$. Using eq.(\ref{|k>}) the latter 
factor is absorbed into $e^{-\frac{i}{4e}\tilde{b} k}$.  
Thus $f_{N}$ are written as (finite) linear combinations 
(and integrations over $k$ and $k'$) of $L^n_1e^{\frac{i}{2e}\tilde{b} k'} G(a, k)$, where 
$G(a, k)$ is a polynomial of $a$.
The explicit expression is 
\begin{eqnarray}
        f_N(x, a) &=& \sum_{n = 0}^{N}\sum_{m = 0}^{n}\sum_{J = 
0}^{n - m}\int_k\int_{k'}\tilde{f}_{nmJ}(k, k')L^m_1e^{-\frac{i}{4e}\tilde{b} k} 
((k' - k)a)^J|0\rangle  ,
\label{AB6}\\
          \tilde{f}_{nmJ}(k, k') &=& \frac{e^{-\frac{i}{4e} k'^2}(
\frac{i}{4e} (k' - k)^2)^{n - m - J}}{(n - m - J)!J!m!}\hat{f}(k, -k').
\label{AB7}
\end{eqnarray}
Finally let us prove that the function $L^n_1e^{-\frac{i}{4e}\tilde{b} k'} G_J(a)|0\rangle$ 
in r.h.s. of eq.(\ref{AB6}), where $G_J$ is an arbitrary 
polynomial of order $J$, belongs to $H_1$, {\it i.e.}, it is 
written as linear combinations (and integrations over $k$) 
of $u_{nJj}(k)$. Then, from eqs.(\ref{AB5}) and (\ref{AB6}), we see 
$f \in \tilde{H}$, which is the desired result.

Since a polynomial $G(a)$, satisfying $
        {\rule[-1mm]{0.1mm}{4mm}
         \rule[3mm]{4mm}{0.1mm} 
         \hspace{-4mm}
         \rule[-1mm]{4mm}{0.1mm} 
         \rule[-1mm]{0.1mm}{4mm}
         \hspace{1mm}}_aG(a) = 0$, is 
written as a linear combination of the harmonic polynomials we 
see, from eq.(\ref{A3}), that,
\begin{eqnarray}
   L_{-1}G_J(a)|0\rangle  = 0  \quad\Longrightarrow  
\quad G_J(a)|0\rangle \in H_1.
\label{AB11}
\end{eqnarray}
Let us prove the fact that there exist coefficients $a_{nJj'}(k)$ 
for any $J$, satisfying
\begin{eqnarray}
  G_J(a)|0\rangle  = \sum_{n = 0}^{[\frac{J}{2}]}\sum_{J' = 
0}^{J}\int_ka_{nJ'j'}(k)u_{nJ'j'}(k),
\label{AB12}
\end{eqnarray}
{\it i.e.}, $G_J|0\rangle \in H_1$.
The proof is given by induction on $J$.
For $J = 1$, $G_J$ is a linear function of $a$ and 
$
        {\rule[-1mm]{0.1mm}{4mm}
         \rule[3mm]{4mm}{0.1mm} 
         \hspace{-4mm}
         \rule[-1mm]{4mm}{0.1mm} 
         \rule[-1mm]{0.1mm}{4mm}
         \hspace{1mm}}_aG_J = 0$, 
hence $L_{-1}G_J|0\rangle  = 0$. Hence by 
(\ref{AB11}) we see $G_J\in H_1$.
Next assume $G_J$ satisfy eq.(\ref{AB12}) for $J \le  J_0$.
By eq.(\ref{A3}) we see
\begin{eqnarray}
         L_{-1}G_{J_0 + 1}|0\rangle  =  -\frac{i}{4e}
        {\rule[-1mm]{0.1mm}{4mm}
         \rule[3mm]{4mm}{0.1mm} 
         \hspace{-4mm}
         \rule[-1mm]{4mm}{0.1mm} 
         \rule[-1mm]{0.1mm}{4mm}
         \hspace{1mm}}_aG_{J_0 
+ 1}|0\rangle .
\label{AB13}
\end{eqnarray}
Since $
        {\rule[-1mm]{0.1mm}{4mm}
         \rule[3mm]{4mm}{0.1mm} 
         \hspace{-4mm}
         \rule[-1mm]{4mm}{0.1mm} 
         \rule[-1mm]{0.1mm}{4mm}
         \hspace{1mm}}_aG_{J_0 + 1}$ is of order $J_0 - 1$ we can use the 
assumption of the induction. Then write
\begin{eqnarray}
        {\rule[-1mm]{0.1mm}{4mm}
         \rule[3mm]{4mm}{0.1mm} 
         \hspace{-4mm}
         \rule[-1mm]{4mm}{0.1mm} 
         \rule[-1mm]{0.1mm}{4mm}
         \hspace{1mm}}_aG_{J_0 + 1}|0\rangle  =  
\sum_{nJj}\int_{k}a_{nJj}(k)u_{nJj}(k).
\label{AB14}
\end{eqnarray}
{}From eqs.(\ref{A5}) and (\ref{A6}) we see
\begin{eqnarray}
    L_{-1}L_1u_{nJj}(k) = (n + 1)(n + \frac{D}{2} + J)u_{nJj}(k).
\label{AB15}
\end{eqnarray}
Then from eqs.(\ref{AB13}), (\ref{AB14}) and (\ref{AB15}) we see
\begin{eqnarray}
    L_{-1}\left( G_{J_0 + 1}  + 
\frac{i}{4e}\sum_{nJj}\int_{k}\frac{a_{nJj}(k)}{(n + 1)(n + 
\frac{D}{2} + J)}u_{nJj}(k)\right) |0\rangle  = 0.
\label{AB16}
\end{eqnarray}
And finally from the fact of (\ref{AB11}) we find the function 
inside the parenthesis of eq.(\ref{AB16}) multiplied by $|0\rangle$, 
and hence $G_{J_0 + 1}|0\rangle$, belong to $H_1$, which completes the 
inductive proof. 
 
\newpage

\newpage
 
\end{normalsize}

\begin{thebibliography}{99}
\bibitem{hori1} T. Hori, J. Phys. Soc. Jap., {\bf 61} (1992) 744.
\bibitem{hori2} T. Hori, Phys. Rev. {\bf D48} (1993) R444.
\bibitem{Dirac} P. A. M. Dirac, Can. J. Math., {\bf 2} (1950) 129.
\bibitem{KO} T. Kugo and I. Ojima, Phys. Lett. {\bf 73B} (1978) 459.
\bibitem{Take} M. Takeuchi,  Modern Spherical Functions (Iwanami 
Shoten, Tokyo, 1975) p.256.
\end{thebibliography}
\end{document}